\documentclass[prb,aps,showpacs,superscriptaddress,twocolumn]{revtex4}

\usepackage{graphicx}

\begin{document}

\title{Certain quantum key distribution achieved by using Bell states}
\author{Chong Li}
\affiliation{Institute of Theoretical Physics, Chinese Academy of
Sciences, Beijing, 100080, China}
\author{He-Shan Song}
\affiliation{Department of physics, Dalian University of
Technology, P.R.China 116024}
\author{Ling Zhou}
\affiliation{Department of physics, Dalian University of
Technology, P.R.China 116024}
\author{Chun-Feng Wu}
\affiliation{Department of Physics, National University of
Singapore, 2 Science Drive 3, Singapore 117542}
\date{\today}

\begin{abstract}
A new protocol for quantum key distribution based on entanglement
swapping is presented. In this protocol, both certain key and
random key can be generated without any loss of security. It is
this property differs our protocol from the previous
ones\cite{14}. The rate of generated key bits per particle is
improved which can approach six bits (4 random bits and 2 certain
bits) per four particles.
\end{abstract}

\maketitle

In cryptography, a message is uneavesdropped to any eavesdroppers.
To achieve this goal, the message is combined with a key to
produce a cryptogram. For a cipher to be secure, it should be
impossible to unlock the cryptogram without the key. In this
sense, the security of the cipher depends on the security of the
key, which is difficult to be assured by classical means. Quantum
cryptography, also called quantum key distribution (QKD), is
defined as a procedure, in which two legitimate communicators can
establish a sequence of key bits secretly and any unauthorized
user can be detected. Since the \textbf{BB84} protocol[1],the
first quantum key distribution scheme, was proposed, various
quantum encryption schemes have been proposed such as \textbf{B92}
protocol[2] , the EPR protocol[3] and other protocols[4-11]. All
these protocols for QKD produce random key bits, the security of
the protocols is assured by this way. Here we propose a new QKD
scheme in which not only random key bits, but also certain key
bits can be generated without any loss of security.

In this paper, we present a new protocol for quantum key
distribution based on entanglement swapping. In our proposed
protocol, \emph{Alice} only transports $2$ particles to
\emph{Bob}, she can share $6$ bits, which include $2$ certain bits
and $4$ random bits, with \emph{Bob} secretly. This protocol
brings some merits over the previous protocols, for example, the
use efficiency can approach $100\%$

\begin{figure}[ht]
\includegraphics [angle=-90, totalheight=5cm]{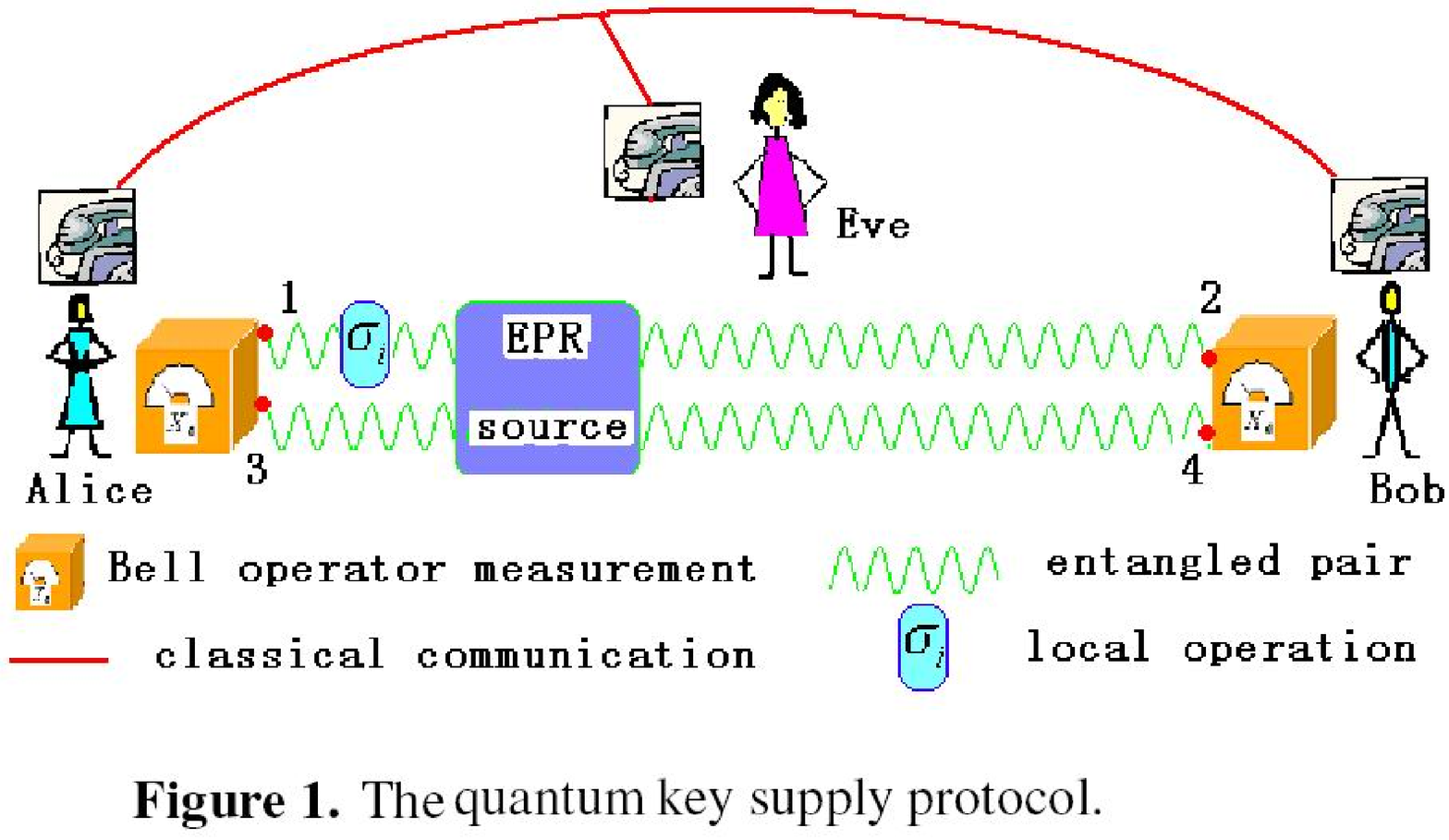}

\end{figure}

To illustrate our new protocol for QKD, we explain the process of
entanglement swapping first.

It has been proposed by Zukowski et al.[13] for two pairs of
entangled particles with each pair in one of the Bell states.
Entanglement swapping works as follows. Consider two pairs of

entangled particles $1$, $2$, $3$ and $4$, prepared in Bell states
respectively $\left\vert \phi ^{+}\right\rangle _{12}$ and
$\left\vert \psi ^{-}\right\rangle _{34}$. If a Bell operator
measurement is performed on particles $2$ and $3$, then we have

\begin{eqnarray}
\left\vert \Psi \right\rangle _{1234} &=&\left\vert \phi
^{+}\right\rangle
_{12}\otimes \left\vert \psi ^{-}\right\rangle _{34}  \nonumber \\
&=&\frac{1}{2}\{\left\vert \phi ^{+}\right\rangle _{23}\left\vert
\psi ^{-}\right\rangle _{14}+\left\vert \phi ^{-}\right\rangle
_{23}\left\vert
\psi ^{+}\right\rangle _{14}  \nonumber \\
&&-\left\vert \psi ^{+}\right\rangle _{23}\left\vert \phi
^{-}\right\rangle _{14}-\left\vert \psi ^{-}\right\rangle
_{23}\left\vert \phi ^{+}\right\rangle _{14}\}
\end{eqnarray}
As is obvious, from above equation, the four possible results
$\left\vert
\phi ^{+}\right\rangle _{23}$,$\left\vert \phi ^{-}\right\rangle _{23}$,$%
\left\vert \psi ^{+}\right\rangle _{23}$ and $\left\vert \psi
^{-}\right\rangle _{23}$ occur with same probability and the
outcome of each measurement is purely random. Suppose that the
result $\left\vert \phi ^{+}\right\rangle _{23}$ is obtained, the
state of the pair $1$ and $4$ after the measurement is
consequently $\left\vert \psi ^{-}\right\rangle _{14}$. Which
means particles $1$ and $4$ become entangled although they have
never interacted.

Recently, the entanglement swapping between two pairs of qubits
has been used to realize schemes for quantum key distribution[9].
Which do not need the legitimate users to choose between possible
measurements to generate the key and assure the security. But only
random key bits can be produced in these schemes.

Our scheme, illustrated in Fig. 1, can be described as follows.

\emph{Alice} has a EPR source. She can apply a local operation $X$ , where $%
X\in \left\{ \sigma _{0}\text{,}\sigma _{1}\text{,}\sigma
_{2}\text{,}\sigma _{3}\right\} $ on her particles. \emph{Alice}
and \emph{Bob} agree beforehand as following encoding

\begin{equation}
\left\vert \phi ^{+}\right\rangle \rightarrow
00\text{,.}\left\vert \psi ^{+}\right\rangle \rightarrow 01\text{,
}\left\vert \psi ^{-}\right\rangle \rightarrow 10\text{,
}\left\vert \phi ^{-}\right\rangle \rightarrow 11
\end{equation}

where$\left\vert \phi ^{\pm }\right\rangle $ and $\left\vert \psi
^{\pm }\right\rangle $ are Bell states.

\begin{equation}
\sigma _{0}\rightarrow 00\text{,}\sigma _{1}\rightarrow
01\text{,}\sigma _{2}\rightarrow 10\text{,}\sigma _{3}\rightarrow
11
\end{equation}

They can perform series of operations as following,

1. \emph{Alice} creates EPR pairs and sends some of her particles to \emph{%
Bob}.

\emph{Alice} creates two EPR pairs $\left\vert \phi
^{+}\right\rangle _{12}$ and $\left\vert \phi ^{+}\right\rangle
_{34}$, she sent particles $2$ and $4$ to

\emph{Bob},and tells \emph{Bob} the state in which the particles
are.

2. \emph{Alice} applies a local operation on particle $1$.

Let the initial state of particles $1$ and $2$ be $\left\vert \phi
^{+}\right\rangle _{AB}^{12}$, and the state of particles $3$ and $4$ be $%
\left\vert \phi ^{+}\right\rangle _{AB}^{34}$. \emph{Alice}
performs a local operation $\sigma _{1}$ on particle $1$ and thus,
the state $\left\vert \phi ^{+}\right\rangle _{AB}^{12}$ is turned
into $\left\vert \psi ^{+}\right\rangle _{AB}^{12}$.

3. \emph{Alice} makes a Bell operator measurement on particles $1$
and $3$. We assume that the result of \emph{Alice}'s measurement
as $\left\vert \psi ^{-}\right\rangle _{AA}^{13}$, then she can
infer the state of particles $2$ and $4$ as $\left\vert \phi
^{-}\right\rangle _{BB}^{24}$ by the following equation,

\begin{eqnarray}
\left\vert \Phi \right\rangle _{ABAB}^{1234} &=&\left\vert \psi
^{+}\right\rangle _{AB}^{12}\otimes \left\vert \phi
^{+}\right\rangle
_{AB}^{34}  \nonumber \\
&=&\frac{1}{2}\{\left\vert \phi ^{+}\right\rangle
_{AA}^{13}\left\vert \psi ^{+}\right\rangle _{BB}^{24}-\left\vert
\phi ^{-}\right\rangle
_{AA}^{13}\left\vert \psi ^{-}\right\rangle _{BB}^{24}  \nonumber \\
&&+\left\vert \psi ^{+}\right\rangle _{AA}^{13}\left\vert \phi
^{+}\right\rangle _{BB}^{24}-\left\vert \psi ^{-}\right\rangle
_{AA}^{13}\left\vert \phi ^{-}\right\rangle _{BB}^{24}\}
\end{eqnarray}

4. \emph{Alice} calculates which state of particles $1$ and $3$
should be without local operation applied on particle $1$ when the
EPR state of particles $2$ and $4$ is $\left\vert \phi
^{-}\right\rangle _{BB}^{24}$.

\begin{eqnarray}
\left\vert \Phi \right\rangle _{ABAB}^{1234} &=&\left\vert \phi
^{+}\right\rangle _{AB}^{12}\otimes \left\vert \phi
^{+}\right\rangle
_{AB}^{34}  \nonumber \\
&=&\frac{1}{2}\{\left\vert \phi ^{+}\right\rangle
_{AA}^{13}\left\vert \phi ^{+}\right\rangle _{BB}^{24}+\left\vert
\phi ^{-}\right\rangle
_{AA}^{13}\left\vert \phi ^{-}\right\rangle _{BB}^{24}  \nonumber \\
&&+\left\vert \psi ^{+}\right\rangle _{AA}^{13}\left\vert \psi
^{+}\right\rangle _{BB}^{24}+\left\vert \psi ^{-}\right\rangle
_{AA}^{13}\left\vert \psi ^{-}\right\rangle _{BB}^{24}\}
\end{eqnarray}

From Eq.($5$), \emph{Alice} get the state of particles $1$ and $3$ as $%
\left\vert \phi ^{-}\right\rangle _{AA}^{13}$, compared this
result with expression($3$), she decodes the bits as $11$.

5. \emph{Alice} tells \emph{Bob} she has made a Bell operator
measurement on her particles $1$ and $3$ ,but without mention the
result of her measurement through classical channel.

6. \emph{Bob} performs a Bell operator measurement on particles
$2$ and $4$ and infers the outcomes of \emph{Alice}'s measurement.

From the calculation of entanglement swapping, we know, after
\emph{Alice} has performed a Bell operator on particles $1$ and
$3$, \emph{Bob}'s measurement result of his particles should be
$\left\vert \phi ^{-}\right\rangle _{BB}^{24}$. Note that the
result of \emph{Alice}'s measurement inferred by \emph{Bob} is the
corresponding state determined using Eq.($5$). Because \emph{Bob}
does not know the exact state of entangled pair $1$ and $2$ after
\emph{Alice} subjects her particle $1$ to a local operation. What
he can do is to use known information: entangled pairs $\left\vert
\phi ^{+}\right\rangle _{AB}^{12}$ and $\left\vert \phi
^{+}\right\rangle _{AB}^{34}$, to calculate the state particles
$1$ and $3$ is in. In this case, it is $\left\vert \phi
^{-}\right\rangle _{AA}^{13}$, which can be decoded as $11$.

 7.\emph{Bob} asks \emph{Alice}'s corresponding outcome of her Bell
operator measurement on particles $1$ and $3$ by classical
channel. He knows the Bell operator measurement of particles $1$
and $3$ is $\left\vert \psi ^{-}\right\rangle _{AA}^{13}$.

8. \emph{Bob} compares the measurement result of particles $1$ and
$3$ announced by \emph{Alice} with his calculation result, and he
will find what the \emph{Alice}'s local operation on particle $1$
is

$\left\vert \psi ^{-}\right\rangle _{AA}^{13}{\sigma _{1}}{
\longleftrightarrow }\left\vert \phi ^{-}\right\rangle _{AA}^{13}$, \emph{Bob%
} gets the local operation is $\sigma _{1}$. Then he knows the
certain bits

is $01$. Obviously, both \emph{Alice} and \emph{Bob} know the state $%
\left\vert \phi ^{-}\right\rangle _{BB}^{24}$ and the imaginary
state of particles $1$ and $3$ as $\left\vert \phi
^{-}\right\rangle _{AA}^{13}$ secretly. Thus they establish the
random key bits $11$ at the same time.

\emph{Bob} makes another ES calculation as eq.($5$), he obtain the
imaginary states of particles $1$ and $3$ as $\left\vert \psi
^{-}\right\rangle _{AA}^{13}$. He decodes the random bits as $10$.

They perform above procedures on each group of particles repeat.

9. \emph{Bob} told \emph{Alice} some his qubits, \emph{Alice} will
be sure if there is any eavesdropper.

Then \emph{Alice} and \emph{Bob} securely share the certain key
bit and the random key.

\section{Security}

Security is an important issue of QKD. One protocol for QKD is
said to be secure if it can either generate the key secretly or
stops the protocol when any eavesdroppers appear. In this section,
we proof that our proposed protocol is secure, even if the shared
quantum channels are public.

First, we consider the security of the random key which is encoded
in the
outcomes of \emph{Bob}'s measurement and the calculation result of "\emph{%
Alice}'s measurement"assuming that she does not apply any local
operations on her particle. Where the symbol ""implies that
\emph{Alice} does not really apply this measurement and get the
outcome, it is just an imaginary
measurement. After \emph{Alice} has performed a local operation on particle $%
1$ and a Bell operator measurement, no eavesdroppers can gain any
information, although \emph{Alice} announces publicly her Bell
measurement
result on particles $1$ and $3$. The reason is that no one except \emph{Alice%
} knows exactly the entangled states of particles $1$ and $2$ once \emph{%
Alice} performs a local operation on her particle. There is only a
probability of $\frac{1}{4}$ to guess the correct EPR state in
which particles $1$ and $2$ are in. If Eve eavesdrops $4n$ key
bits , she tends to succeed with a probability of $\left(
\frac{1}{4}\right) ^{n}$ which approaches $0$ and $n$ is
sufficiently large. Further, if Eve is so clever that she shares
the quantum channels with \emph{Alice} and \emph{Bob}, or replaces
the particles in transit to \emph{Alice} and \emph{Bob} by the
particles prepared by her, she can succeed in eavesdropping with
probability $\frac{1}{2}$ or $\frac{1}{4}$respectively. This has
been certified by us in[14]. Then \emph{Bob} can send some results
to \emph{Alice} to detect the action of Eve and thus, Eve cannot
gain any information of the random key.

Now turns to the problem of the certain key which is encoded in
the local operations performed by \emph{Alice} on her particle. It
is available to
\emph{Bob} by comparing the state of particles $1$ and $3$ gotten by \emph{%
Bob} with the result announced by \emph{Alice}. Thus the security
of the certain key depends on whether \emph{Bob} can secretly
infer the correct corresponding state of particles $1$ and $3$
using the entanglement swapping between two initial shared
entangled pairs. In our previous work[14], we showed that it is
impossible for any eavesdroppers to know the Bell operator
measurement results by innerspring the transmission. More detail,
Eve cannot determine the outcomes of the sender and receiver's
measurements in which the key is encoded, no matter what methods
Eve may choose. There is a probability of $\frac{1}{2}$ to gain
the correct information on the key when Eve tries to share the
quantum communication channels with \emph{Alice} and \emph{Bob},
and a success probability of $\frac{1}{4}$ by replacing the
particles transported to legitimate users by the particles
accessible to her. This is a useful conclusion even for this
different protocol. That is to say, Eve can eavesdrop no
information of the certain key.

This protocol is secure, even the shared entangled states are
public (in another words, everyone know that the \emph{Alice} and
\emph{Bob} shared states are $\left\vert \phi ^{+}\right\rangle
_{AB}$ ). If \emph{Alice} and \emph{Bob} do not publish their's
measured result, eavesdropper (\emph{Eve}) can not get the
information from the known entangled channels. For example, both
the quantum channels are the same \emph{Bell} states $\left\vert
\phi ^{+}\right\rangle _{AB}^{12}$ and $\left\vert \phi
^{+}\right\rangle _{AB}^{34}$. By the entanglement swapping
calculation, we know the state of these entangled particles are

\begin{eqnarray}
\left\vert \Phi \right\rangle _{ABAB}^{1234}
&=&\frac{1}{2}\{\left\vert \phi ^{+}\right\rangle
_{AA}^{13}\left\vert \phi ^{+}\right\rangle _{BB}^{24}+\left\vert
\phi ^{-}\right\rangle _{AA}^{13}\left\vert \phi
^{-}\right\rangle _{BB}^{24}  \nonumber \\
&&+\left\vert \psi ^{+}\right\rangle _{AA}^{13}\left\vert \psi
^{+}\right\rangle _{BB}^{24}+\left\vert \psi ^{-}\right\rangle
_{AA}^{13}\left\vert \psi ^{-}\right\rangle _{BB}^{24}.
\end{eqnarray}

So the four Bell-operator measurement outcomes of Alice are
equally likely, each occurring with probability $\frac{1}{4}$. \
Yet the eavesdropper can
not get any information, although she owned many entangled pairs in state $%
\left| \phi ^{+}\right\rangle _{EE}$.

If the \emph{Eve} was smart enough, she made the state $\left|
\phi ^{+}\right\rangle _{ABE}^{\prime }=\frac{1}{\sqrt{2}}\left(
\left| 00\right\rangle _{AB}\left| \alpha \right\rangle
_{E}+\left| 11\right\rangle _{AB}\left| \beta \right\rangle
_{E}\right) $ instead of $\left| \phi ^{+}\right\rangle _{AB}$,
when \emph{Alice} made a joint measurement on particle $1$ and
particle $3$, the state become

\begin{eqnarray}
&&\left\vert \Psi \right\rangle _{ABE}^{123456}  \nonumber \\
&=&\left\vert \phi ^{+}\right\rangle _{ABE}^{\prime }\otimes
\left\vert \phi
^{+}\right\rangle _{ABE}^{\prime }  \nonumber \\
&=&\frac{1}{2\sqrt{2}}\{\left\vert \phi ^{+}\right\rangle
_{AA}^{13}(\left\vert \phi ^{+}\right\rangle _{BB}^{24}\left\vert
\phi ^{+}\right\rangle _{EE}^{56}+\left\vert \phi
^{-}\right\rangle
_{BB}^{24}\left\vert \phi ^{-}\right\rangle _{EE}^{56})  \nonumber \\
&&+(\left\vert \phi ^{-}\right\rangle _{AA}^{13}\left\vert \phi
^{+}\right\rangle _{BB}^{24}\left\vert \phi ^{-}\right\rangle
_{EE}^{56}+\left\vert \phi ^{-}\right\rangle _{BB}^{24}\left\vert
\phi
^{+}\right\rangle _{EE}^{56})  \nonumber \\
&&+(\left\vert \psi ^{+}\right\rangle _{AA}^{13}\left\vert \psi
^{+}\right\rangle _{BB}^{24}\left\vert \psi ^{+}\right\rangle
_{EE}^{56}+\left\vert \psi ^{-}\right\rangle _{BB}^{24}\left\vert
\psi
^{-}\right\rangle _{EE}^{56})  \nonumber \\
&&+\left\vert \psi ^{-}\right\rangle _{AA}^{13}(\left\vert \psi
^{+}\right\rangle _{BB}^{24}\left\vert \psi ^{-}\right\rangle
_{EE}^{56}+\left\vert \psi ^{-}\right\rangle _{BB}^{24}\left\vert
\psi ^{+}\right\rangle _{EE}^{56})\}.
\end{eqnarray}

Compare eq.[$3$] with eq.[$4$] , we found if there is an
eavesdropper, there is only $\frac{1}{2}$ probability that the two
result are same. \emph{Bob} can send some result random to
\emph{Alice}, then \emph{Alice} can know there is eavesdropper
with different result, they give up this key. Namely, \emph{Eve}
can not gain any information about the key.

If \emph{Eve }shared the entangled pairs $\left\vert \phi
^{+}\right\rangle _{AE}^{ij}$ with \emph{Alice} instead of
\emph{Bob}, \emph{Eve }shared entangled pairs $\left\vert \phi
^{+}\right\rangle _{EB}^{i^{/}j^{/}}$ with \emph{Bob}, before the
key distribution, both \emph{Alice} and \emph{Bob} do not know
this. \ Then the process of key distribution become\

\begin{eqnarray}
\left\vert \Phi \right\rangle _{AEAE}^{1234}
&=&\frac{1}{2}\{\left\vert \phi ^{+}\right\rangle
_{AA}^{13}\left\vert \phi ^{+}\right\rangle _{EE}^{24}+\left\vert
\phi ^{-}\right\rangle _{AA}^{13}\left\vert \phi
^{-}\right\rangle _{EE}^{24}  \nonumber \\
&&+\left\vert \psi ^{+}\right\rangle _{AA}^{13}\left\vert \psi
^{+}\right\rangle _{EE}^{24}+\left\vert \psi ^{-}\right\rangle
_{AA}^{13}\left\vert \psi ^{-}\right\rangle _{EE}^{24}\}
\end{eqnarray}

and

\begin{eqnarray}
\left\vert \Phi ^{/}\right\rangle _{EBEB}^{1^{/}2^{/}3^{/}4^{/}} &=&\frac{1}{%
2}\{\left\vert \phi ^{+}\right\rangle _{EE}^{1^{/}3^{/}}\left\vert
\phi ^{+}\right\rangle _{BB}^{2^{/}4^{/}}+\left\vert \phi
^{-}\right\rangle _{EE}^{1^{/}3^{/}}\left\vert \phi
^{-}\right\rangle _{BB}^{2^{/}4^{/}}
\nonumber \\
&&+\left\vert \psi ^{+}\right\rangle _{EE}^{1^{/}3^{/}}\left\vert
\psi ^{+}\right\rangle _{BB}^{2^{/}4^{/}}+\left\vert \psi
^{-}\right\rangle _{EE}^{1^{/}3^{/}}\left\vert \psi
^{-}\right\rangle _{BB}^{2^{/}4^{/}}\}
\end{eqnarray}
From Eq.($5$) and Eq.($6$), we know that, the probability, which \emph{Bell }%
operator measurement of particle $1$ and $3$ is same as particle
$1^{/}$ and $3^{/}$, is only $\frac{1}{4}$, when Bob send some his
measurement results to Alice, Alice can find if there is
eavesdropper.

So this protocol is secret and secure.

\begin{acknowledgments}
This work was partially supported by the CNSF (grant No.90203018)
,the Knowledged Innovation Program (KIP) of the Chinese Academy of
Sciences , the National Fundamental Research Program of China with
No.001GB309310, K. C. Wong Education Foundation, HongKong, and
China Postdoctoral Science Foundation.
\end{acknowledgments}


\begin{thebibliography}{DiVincenzo(2000)}
\bibitem{1} C.H.Bennett, G.Brassard, in proc IEEE int conf on Computers
System,and Signal Processing, Bangalore (IEEE, New York, 1984)P175

\bibitem{2} C.H.Bennett, Phys.Rev.Lett \ \textbf{68} (1992)3121

\bibitem{3} A.K.Ekert, Phys.Rev.Lett \textbf{67} (1991)661

\bibitem{4} C. H. Bennett, G. Brassard and N. D. Mermin, Phys.Rev. Lett.
\textbf{68} (1992) 557.

\bibitem{5} B. Huttner, N. Imoto, N. Gisin and T. Mor, Phys. Rev.\textbf{A 51%
} (1995) 1863.

\bibitem{6} D. Bruss , Phys. Rev. Lett. \textbf{81}(1998) 3018.

\bibitem{7} L. Goldenberg and L. Vaidman, Phys. Rev. Lett. \textbf{75}
(1995) 1239

\bibitem{8} M. Koashi and N. Imoto, Phys. Rev. Lett. \textbf{79} (1997)2383.

\bibitem{9} A.CaBelloperatoro, Phys. Rev. \textbf{A 61} (2000) 052312

\bibitem{10} G-C Guo and B-S Shi, Phys. Lett. \textbf{A, 256},109 (1999)

\bibitem{11} B.-S Shi, Y.-K Jiang and G.-C Guo, Appl. Phys. \textbf{B, 70}%
,415, (2000)

\bibitem{12} G. L. Long and X.S. Liu, quant-ph/0012056

\bibitem{13} M. Zukowski, A. Zeilinger, M. A. Horne, and A. K.Ekert,Phys.
Rev. Lett \textbf{71}, 4287 (1993).

\bibitem{14} C. Li, H.S. Song, L. Zhou and C.F. Wu, J. of Opt. B: Quantum
and semi-classical optics \textbf{5} (2003)155
\end{thebibliography}
\end{document}